# Beam quality measurement through off-axis optical vortex

August 2019

**Authors:**

Mateusz Szatkowski*[a], Agnieszka Popiołek - Masajada[a], Jan Masajada[a]

**Addresses:**

[a] Department of Optics and Photonics, Faculty of Fundamental Problems of Technology, Wroclaw University of Science and Technology, Wyb. Wyspianskiego 27, 50-370 Wroclaw, Poland

*mateusz.szatkowski@pwr.edu.pl

**ABSTRACT**

One of the challenges for every optical system is preserving the quality of the used beam, which may be significantly reduced, due to the low condition of used optical elements or their misalignment. There are plenty of methods focused on correction of the final beam, depending on the used entire optical system. Problem of efficient beam evaluation is just as important. So far, most of the measurements, are based on visual inspection, which is not always enough, especially when the high quality of the beam is required. Novel approaches use structured light to increase beam sensitivity for any imperfections. In this paper we present approach, which uses optical vortex shifted off-axis for a beam quality measurement. It uses SLM as a vortex generating element, which is shifted off-axis by proper hologram transformation. Tracking of the vortex trajectory may provide information about beam quality and aberration of optical system. The new vortex localization algorithm will be presented.

1. **INTRODUCTION**

Together with the broadening popularity of laser beam shaping, the demand on high quality spatial light modulators (SLM) is simultaneously increasing. Some imperfections of these devices can be overwhelmed. Nevertheless, none of correction methods can work without proper beam evaluation. The typical approach to beam evaluation is based on visual inspection of corrected beam. First, the most common approach is an inspection of Gaussian beam through the Strehl ratio[1]. Most of recent approach uses the structured light for these purposes, where laser beam can be shaped into almost any desired output. One of such examples examines the beam quality through optical vortex image, taking the advantage of reach internal structure of optical vortex. The phase gradient of optical vortex changes by integer number of 2π around a singular point. This local phase gradient exceeds the wavenumber $k = 2\pi/\lambda$ and goes to infinity. Such quick variations are called superoscillations. This phenomenon refers to all signals, which locally oscillate quicker than its global Fourier components[2].

Optical vortex image criterion can be combined with Gerchberg – Saxton (G-S) phase retrieval algorithm for SLM correction purposes[3]. This method uses optical vortex as a beam quality marker. The algorithm searches for SLM correction phase map, which will transform the imperfect optical vortex into perfect one. In the algorithm, the process is reverted, it searches for the phase map that produces disturbed vortex (registered experimentally) and then by subtraction of retrieved phase map from final hologram, the imperfections of the whole optical system are overwhelmed. The aspect of proper criterion, in case of G-S algorithm is crucial. It is based on the comparison between experimental and simulated optical vortex image, the algorithm stops, when the global minimum is reached. Another, more objective approach was proposed together with SLM correction using Zernike polynomials[4] in the optical trapping setup. The evaluation of final beam was based on tracing the symmetry of trapped polystyrene microbeads positions. At the end it was reduced to examination of trap stiffness for both Gaussian and Laguerre-Gaussian modes, where more symmetric beam provided higher stiffness. This, more objective method of beam evaluation is limited to optical trapping setups and requires additional elements to perform (such as polystyrene microbeads). Recently, the method of beam quality examination through optical vortex dynamics was presented[5]. Optical vortex dynamics is expressed as an optical vortex trajectory[6], detected at observation plane, when vortex is shifted off-axis. In the perfect optical system, vortex travels along the straight line. Examination of this optical vortex trajectory is widely explored in the Optical Vortex Scanning Microscope (OVSM) project, where its shape is one of the possibilities to obtain information about the measured object[7]. So far, OVSM uses spiral phase plate as a vortex generating element, mainly due to relatively good quality of produced optical vortex. Reproduction of a straight vortex trajectory, with SLM as a generating element, became an obvious quality criterion in order to exchange spiral phase plate with SLM, as a vortex generator. This could open a new possibility for the OVSM, offering a higher dynamic. Simultaneously, it can act as an objective and precise method to evaluate beam quality. In this paper the new methodology for localization of optical vortex on registered image will be presented. This methodology uses Laguerre-Gaussian filter to transform detected intensity signal to pseudo-complex signal[8], on which the vortex point can be easily localized.

## 2. OFF-AXIS OPTICAL VORTEX AND ITS TRAJECTORY

Gaussian beam, aligned with vortex generating element produces a symmetric intensity donut shape with central singular point, where phase is undetermined[9]. When any misalignment between beam and vortex generating element appears, this symmetry is broken (Figure 1).

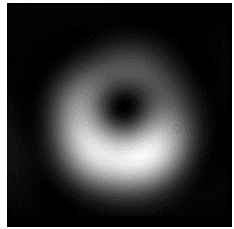

Figure 1. Optical vortex intensity image (experimental), when off-axis shift occurs.

The general idea of optical vortex off-axis shifting is presented in Figure 2, where vortex generating element is translated. However, it requires further comment. The vortex (singular) point travels along a straight line due to vortex generating element movement. This movement, when registered at observation plane is called a vortex trajectory. The angle $\alpha$ between vortex trajectory and direction of element movement depends on the position of the observation plane along the optical axis[10].

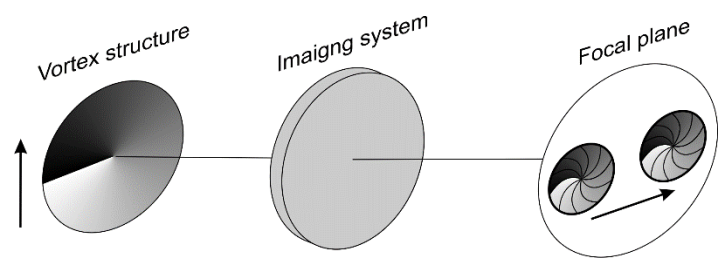

Figure 2. Scheme of optical vortex movement within the beam, due to optical vortex generating element shift.

Close to the Fourier plane, the angle $a = \pi/2$ which means that trajectory is perpendicular to the movement of the generating element. This plane is defined as a critical plane. This optical vortex feature is widely explored in the OVSM[10,11], where shape of this trajectory is one of the channel providing information about a measured object.

In order to examine vortex trajectory with optical vortex generated through SLM, the proper phase structure, must be translated by mathematical operation. This phase structure usually contains diffraction grating together with spiral phase (added modulo 2π) responsible for introducing vortex into 1st order of diffraction. The scheme of this procedure is presented in Figure 3, where phase structure displayed on SLM was shifted in two directions, X and Y. As the result, the vortex point moved within the beam, which is well represented by the movement of dark point on intensity images. The presented experimental images were registered close to the critical plane, thus movement of the vortex point is perpendicular to the vortex structure shift. Results of one of the early experiments which explored vortex trajectory for the vortex generated by SLM were presented in[6]. Unfortunately, the condition of perpendicular trajectory was not satisfied, probably due to insufficient SLM correction. It is important to note, that exploration of a vortex trajectory is possible in a limited range of the beam diameter. If the range of a scan is larger, the vortex moves rapidly towards the edge of the beam and singular point disappears in a dark beam area.

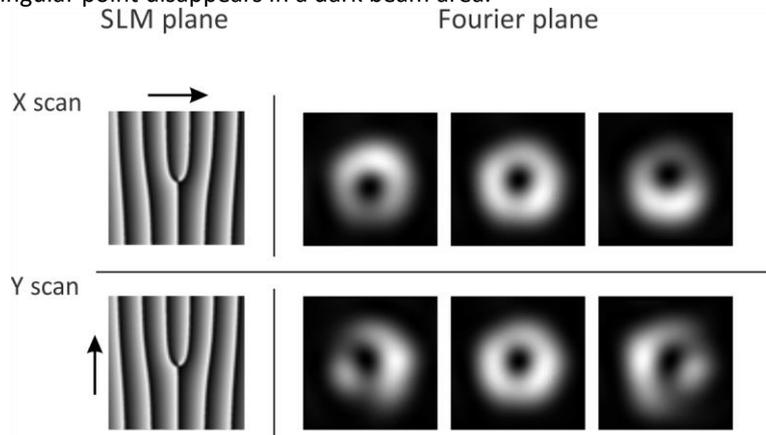

Figure 3. Vortex movement inside the beam, due to translation of vortex generating phase structure (x and y scan).

## 3. EXPERIMENT AND VORTEX LOCALIZATION PROCEDURE

The experimental setup is presented in Figure 4, where expanded laser beam illuminates SLM. Display of proper phase structure with additional diffraction grating allows to direct the desired beam into 1st order of diffraction. Optical vortex is imaged on CCD camera by 4f imaging system. Polarizer together with half wave plate is responsible for polarization control, to obtain efficient 2π phase modulation through SLM. The beam quality evaluation is based on the vortex trajectory inspection. The reproduction of proper shape of this trajectory is a primary criterion for SLM application in OVSM, but more importantly, any trajectory deviation can provide information about optical system imperfections.

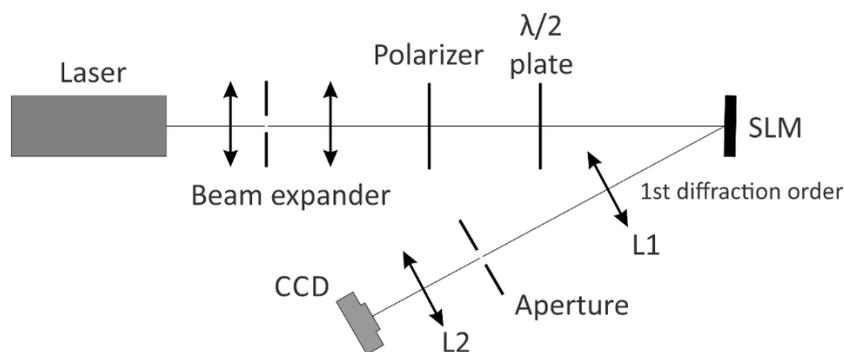

Figure 4. Scheme of the experimental setup.

SLM correction was divided into 2 steps. First step was based on the interferometric measurement of the SLM surface, then in this specific setup, the final vortex image was improved by G-S phase retrieval algorithm[3]. So far, this combination provided the best quality of vortex image, according to our experience. Vortex image was registered on CCD camera. To measure vortex movement inside the beam, the vortex point must be localized on the intensity image. The new method that was adapted for this purposes, uses Laguerre-Gaussian transform[12]. It was previously applied to speckle metrology, revealing speckle-vortex core structure for small displacement tracking[13,14]. The potential to reveal core structure of optical vortex can lead to fast and efficient way to localize the vortex point. The relation between detected intensity signal $I(x,y)$ and its complex analytic signal $\tilde{I}(x,y)$ is represented in the following form:

$$\tilde{I}(x,y) = I(x,y) * \mathbf{LG}(x,y) \qquad (1)$$

The Laguerre-Gaussian filter in Fourier domain is expressed as:

$$LG(f_x, f_y) = \rho \, exp(-\rho^2/\omega^2) exp(j\beta) \qquad (2)$$

where $\rho = sqrt(x^2 + y^2)$, $\beta = \arctan(f_x/f_y)$ denote polar coordinates in the spatial frequency domain and bandwidth $\omega$ is responsible for adjusting $LG(f_x, f_y)$ to match the size of the optical vortex. Following the relation:

$$\begin{aligned}\mathbf{LG}(x,y) = F^{-1}\{LG(f_x, f_y)\} &= (j\pi^2\omega^4)(x+jy)exp(-\pi^2\omega^2(x^2+y^2)) \\ &= (j\pi^2\omega^4)(r\, exp(-\pi^2 r^2 \omega^2) exp(j\alpha))\end{aligned} \qquad (3)$$

where $F^{-1}$ denotes the inverse Fourier transform, and: $r = \sqrt{x^2+y^2}$ together with $\alpha = arctan(x/y)$ represent the spatial polar coordinates. It is important to note, that obtained $\tilde{I}(x,y)$ is pseudo-complex amplitude, however it may provide the valuable information about measured signal. The real and imaginary part, in the close vicinity of the vortex center can be approximated with:

$$Re[\tilde{I}(x,y)] = a_r x + b_r y + c_r, \qquad Im[\tilde{I}(x,y)] = a_i x + b_i y + c_i \qquad (4)$$

Where surface coefficients are obtained by the least-square approximation and (x, y) denote complex values of $\tilde{I}(x,y)$. From both $Re[\tilde{I}(x,y)]$ and $Im[\tilde{I}(x,y)]$, the isolines at value 0 are extrapolated:

$$g_1(x,y) = Re[\tilde{I}(x,y)] = 0, \qquad g_2(x,y) = Im[\tilde{I}(x,y)] = 0 \qquad (5)$$

Potential vortex points can be retrieved by finding the intersection points between these two:

$$g_1(x,y) = g_2(x,y) \qquad (6)$$

Depending on the proper bandwidth $\omega$ selection, more than one potential vortex point can be localized. Condition (6) is presented visually in Figure 5, where cross section of isolines determines the potential vortex points. The final criterion to locate vortex point is to look for one with the lowest intensity. For more noisy images this can come after tight clipping of vortex bright ring with the support of numerical image processing.

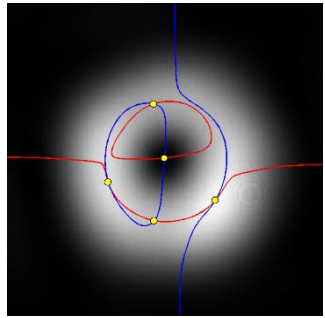

Figure 5. Potential vortex points (yellow dots) determined according to the equation (6), Isoline $g_1(x,y)$ - red line, isoline $g_2(x,y)$ - blue line. The experimental intensity image $I(x,y)$ is used as a background.

The MATLAB code with the implemented vortex localization algorithm can be find in GitHub repository[15], together with the intersection function[16]. The exemplary images, containing localized vortex point, are presented in Figure 6. The presented method provides more accurate vortex positions (red dots), than center of the mass method, used before (yellow crosses). Especially for a scenario where vortex point is placed close to the edge of the beam Figure 6c-d. The vortex point disappears in the outer beam area and the center of mass method, used so far, introduces additional error to the vortex point position. For the discussed case (Figure 6c-d), the difference between vortex positions detected by these two methods was equal to 35 [px]. However, it should be noted, that it is possible to obtain correct vortex position using the center of mass method. It demands intensity limit value adjustment, which defines the range of the points taken into calculation. Nevertheless, it does not allow to analyze all vortex images using the same parameters, which affects the time of analysis and makes automation of this process hardly achievable. This is not true for the new method presented here, where Laguerre-Gaussian transform provides more accurate positions for whole analyzed data set, when bandwidth ω is well determined.

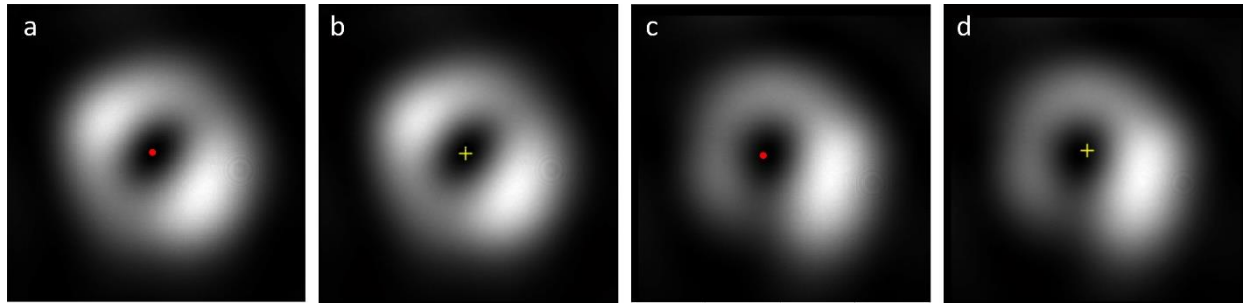

Figure 6. Exemplary experimental images with localized vortex point. a) Optical vortex placed in the center of the beam, vortex point (red dot) was located using L-G transform b) The same, but vortex point (yellow cross) was located using the center of mass method c) Optical vortex placed close to the edge of the beam, vortex point (red dot) was located using L-G method d) The same, but vortex point (yellow cross) was located using the center of mass method.

The performance this new algorithm, was tested in the following experiment. Two (x and y) scans where performed, where vortex phase structure was moved along proper axis of SLM display (Figure 3). These scans were repeated, for the system with manually introduced astigmatism, characterized by Zernike polynomial coefficient defined on the disk having radius $R = 4.316\ mm$:

$$Z_2^{-2} = r^2 \cos 2\theta \qquad (7)$$

This polynomial was combined with the phase structure displayed on SLM. At each step, the vortex point was localized using algorithm described above. Vortex generating structure was shifted at SLM plane in [-2.4 mm, 2.4 mm] range, with 0.018 [mm] step and 0 being central position. The system without manually implemented astigmatism produced two almost straight trajectories, which preserved their perpendicularity in the close vicinity of vortex central position (Figure 7a). In the perfect case those trajectories will resemble straight lines, orthogonal to each other. However, even with advanced SLM correction, both trajectories achieved some curviness at their edges.

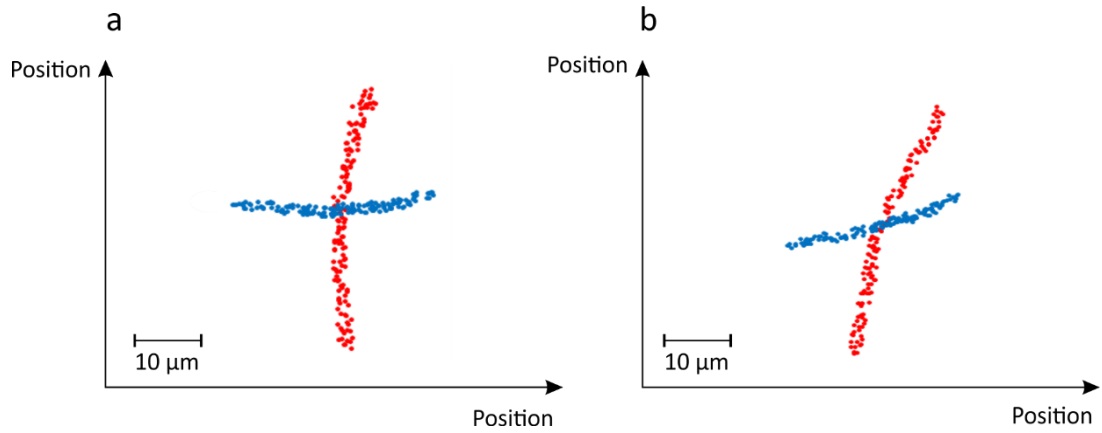

Figure 7. Experimental trajectories obtained for both x (red) and y (blue) scans for the system a) without astigmatism and b) with astigmatism $Z_2^{-2} = 0.54\lambda$.

Second part of the experiment required manually introduced astigmatism. This was done by addition of proper Zernike polynomial to the hologram displayed on SLM. Introduced astigmatism $Z_2^{-2} = 0.54\lambda$ led to inclination of both trajectories (Figure 7b), as was expected. According to the nature of astigmatism, it will have a major impact on registered vortex trajectory. The off-axis vortex shifting will reveal change of optical system foci for sagittal and transverse plane. The fixed observation plane position will lead to inclination of each vortex trajectory. The shape of the trajectory may differ, when the lateral displacement of the vortex from optical axis will be increased.

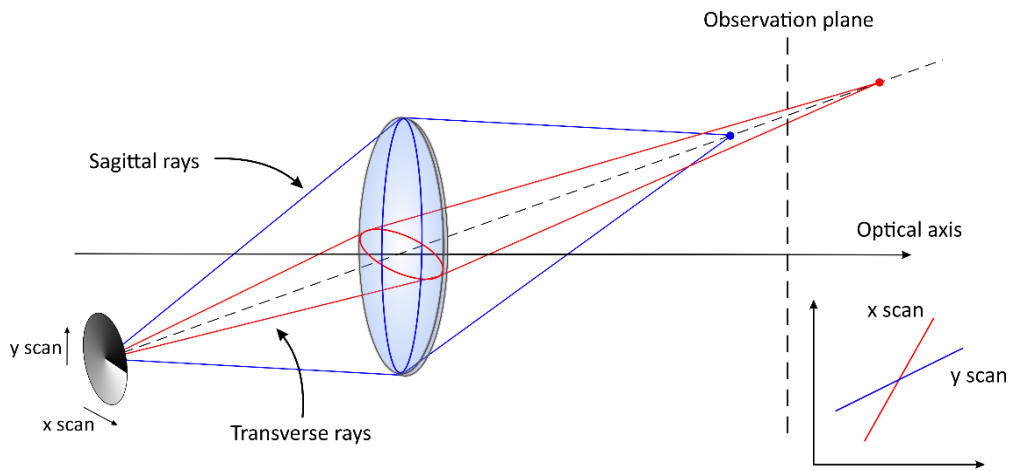

Figure 8. Visualization of optical vortex trajectories in the optical system with astigmatism

The visualization of this behavior is presented in Figure 8. Vortex scan in x direction refers to vortex point movement in the transverse plane (red color). The foci for transverse rays is placed at some distance from an observation plane (dashed line) in the direction of optical axis. Thus, the relation between focal and observation plane position will provide an inclined trajectory. The similar effect will be visible for the y scan in the sagittal plane (blue color). Opposite position of foci (for sagittal rays), provides opposite inclination of the vortex trajectory, if the observation plane position is fixed.

## 4. CONCLUSIONS

The one of the main purposes for efficient beam evaluation method should be its simplicity, which makes it useful for a wide range of applications. The idea standing behind presented approach assumed that this method will be easily adaptive to most of typical optical systems. That was the reason to inspect vortex quality without interference, which could provide more detailed information about vortex phase structure and thus make vortex point localization more convenient. On the other hand, it will limit the range of possible applications. Localization algorithm based on Laguerre-Gaussian transform, offers more sophisticated approach in this matter, comparing to the center of mass method. Additionally, this method may stand itself as a separate, objective way to evaluate beam through the shape of a vortex core structure[5].

Inspection of optical vortex trajectory may provide a direct evaluation of the beam possible. The position of the off-axis vortex and by that, its trajectory, can be affected by even small misalignment or imperfections of an optical system. Our results show that there is still a space for improvement, as can be seen on curved trajectories for the highly corrected beam. Astigmatism was introduced for illustrative purposes, proving that vortex trajectory inspection may possess ability to reveal information about existing aberration. Determination of system aberration from vortex trajectories stands as a separate and due to the complex vortex nature, demanding task. The further research focused on that problem has to be continued.

## 5. ACKNOWLEDGEMENTS

Polish Ministry of Science and Higher Education (" Diamond Grant") (DIA 2016 0079 45);
Nacional Science Centre (Poland) (UMO-2018/28/T/ST2/00125);
Polish Ministry of Science and Higher Education through resources granted to Faculty of Fundamental Problems of Technology in 2018/2019, to support young scientists research and scientific involvement.